\documentclass[11pt]{article}
\usepackage{geometry,amsmath,bm,natbib}
\usepackage{graphicx}
\usepackage{amssymb}
\usepackage[rightcaption]{sidecap}
\newcommand\keywords[1]{\vspace{.1in}\par\noindent{\bf Keywords}: {#1}}
\newcommand\amscode[1]{\vspace{.2in}\par\noindent{\bf 2010 AMS Subject Classification}: {#1}}

\def\la{\lambda}

\def\th{\theta}
\def\bt{\bm{\theta}}
\def\bth{\bm{\theta}}

\def\Q{{\cal Q}}
\def\D{{\cal D}}
\def\d{{\cal D}}

\numberwithin{equation}{section}
\long\def\symbolfootnote[#1]#2{\begingroup
\def\thefootnote{\fnsymbol{footnote}}\footnote[#1]{#2}\endgroup}

\title{Model for Diversity  Analysis of  Antigen Receptor Repertoires}
\author{Grzegorz A. Rempala$^\ast$, Micha{\l} Seweryn$^{\dagger}$, and Leszek Ignatowicz$^\ddagger$}
\date{February 24, 2010}                                        

\begin{document}
\maketitle

\abstract{In  modern  molecular biology  one of the most common ways of  studying  a vertebrate immune system is  to  statistically compare  the   counts of sequenced antigen receptor clones  (either immunoglobulins or T-cell receptors) derived from various  tissues under different   experimental or clinical  conditions. The problem is  difficult  and   does not fit readily into  the standard statistical framework  of contingency tables primarily due to serious under-sampling of the receptor populations. This under-sampling is caused  on one hand  by  the extreme diversity of antigen receptor repertoires maintained by the immune system and, on the other, by  the high cost and labor intensity of   the receptor data collection process.    In most of the recent  immunological literature  the  differences across antigen receptor populations are examined  via  non-parametric statistical measures  of  species overlap  and diversity borrowed from ecological studies.    While this  approach  is     robust in a wide range of situations,  it seems to  provide little insight into  the underlying clonal size distribution and the overall mechanism differentiating the receptor populations.   As a possible alternative,  the current paper presents  a  parametric method   which adjusts for the data under-sampling  as well as  provides a unifying  approach to simultaneous  comparison of  multiple receptor groups by means of the modern statistical tools of unsupervised learning.   The parametric model   is  based  on a  flexible  multivariate Poisson-lognormal distribution   and is seen to be a  natural generalization of the univariate Poisson-lognormal models used in ecological studies of  biodiversity patterns.  The procedure for evaluating  model's fit is described along with the  public domain software  developed to perform the necessary diagnostics.   The  model-driven   analysis  is  seen to compare favorably   vis a vis traditional methods when applied to    the data  from  T-cell receptors in transgenic   mice populations.}
\amscode{62P10, 92B05}

\symbolfootnote[0]{$^\ast$ Corresponding author. Department of Biostatistics and the Cancer Center, Medical College of Georgia, Augusta, GA 30912.   E-mail:grempala@mcg.edu}
\symbolfootnote[0]{$^\dagger$ Wydzia{\l} Matematyki, Universytet {\L}\'odzki, {\L}\'odz, Poland. E-mail msewery@math.uni.lodz.pl}
\symbolfootnote[0]{$^\ddagger$ Department of Medicine, Center for Biotechnology and Genomic Medicine, Medical College of Georgia. E-mail:  lignatowicz@mcg.edu}

\section{Introduction}

The major feature of the adaptive immune system is its  capacity to generate clones of B and T-cells that are able to recognize and neutralize specific antigens. Both cell types recognize antigens by a special class of surface molecules called B- and T-cell receptors. The methodology developed in this paper  will apply to both types of  receptors,  for the sake of clarity and simplicity, we describe the background and the overall problem in terms of  T-cell  receptors.  For a general  introduction to  the molecular biology of the immune system,  we refer  interested reader to e.g.,  \citet{Jane05}. 

A single  T-cell receptor (TCR) is composed of two chains, $\alpha$ and $\beta$, that are formed during T-cell differentiation. Both chains are formed by rearrangements of genetic segments, V$\alpha$ and J$\alpha$ for TCR$\alpha$ chain and V$\beta$, D$\beta$ and J$\beta$ for TCR$\beta$ chain. Since there are  a number of segments of each type in the genomic DNA , a great number of different $\alpha$ and $\beta$ chains are generated. This  chain diversity is further increased by the recombination process when individual nucleotides might be added or deleted at the junctional sites. The region containing these highly variable junctions is the third of three complementarity-determining regions (CDRs) that are seen crystallographically to contact antigen. The sources of TCR diversity are thus naturally broken down hierarchically into gene segment family (library), segment within family, CDR3 length and CDR3 nucleotide diversity.  Both combinatorial and insertional re-arrangements result in the huge TCR repertoire  ensuring that immune system has a potential to recognize a large number of antigens. For instance, it is estimated that in mice the number of different TCRs that can be formed exceeds  $10^{15}$  \citep{Davis:1988fe,Casrouge:2000be}. For humans, it is estimated that over $10^{18}$ different TCRs can be produced and the number of different TCR species ({\em TCR richness})  in a human at any given time has been estimated to exceed ${10^7}$ \citep{Arstila:1999rb,KeithNaylor06012005}. This {\em clonal diversity} of  TCR populations makes them particularly challenging  objects to analyze statistically. 

 In what follows, we  are concerned with  the statistical analysis of the diversity of TCR repertoire samples  obtained from various subsets of T-cells as {\em  counts}  of different TCR clones. These  T-cell subsets are generally defined based on the expression of cell surface markers leading to  different functions in the immune response. For example, naive T-cells are defined as cells that did not encounter an antigen in their lifetime,  while memory T-cells are cells that previously responded to an antigenic stimulation and underwent clonal expansion. The frequency of individual T-cell clones in normal individuals is very low. However, once a naive T-cell expressing appropriate TCR encounters antigen, it becomes activated and expands forming multiple  clone cells.  Thus, the analysis of TCR repertoire  {\em clonal size distribution} has become a crucial element in many studies aimed at better understanding the evolution of the immune responses in vaccinated individuals or patients suffering from autoimmune diseases or cancer  (cf., e.g., \citealt{Butz:1998qt}). The  clonal size estimates has been recently  applied to quantify differences in  TCR repertoires of  normal and infected individuals and to help determine which T-cell clones persisted  as memory T-cells (\citealt{McHeyzer-Williams:1999gt,Sebzda:1999iz, Busch:1999ts,Savage:1999no}). The comparisons of the diversity of TCR repertoires has been  also used to determine the origin of T-cells and  to study heterogeneity of memory T-cells (\citealt{Sallusto:1999rv,Sallusto:2004on}).  For the purpose of the current paper,   under  the  term  ``clonal  diversity" we will understand both the TCRs clonal size distribution (abundance pattern) as well as their richness (number of different TCRs).

There are several statistical approaches to assess T-cell diversity based on  the TCR  clones counts. For instance, in the non-parametric approach, Simpson's diversity and Shannon's entropy indices  have been applied to  measure the diversity of collected  samples \citep{Ferreira:2009fk,Venturi:2008uq}. However, such  indexes are typically just summary statistics and,   due to frequent data under-sampling,  provide only limited information about the   true diversities of the populations.  Another approach seen in immunological studies relies  on modeling  diversity parametrically  assuming that all clonotypes are equally represented in the repertoire (\citealt{Barth:1985kx,Behlke:1985vn}). The advantage of this homogenous model is  its computational and conceptual simplicity  which contributes to its  wide use in analyzing TCR data  (e.g., \citealt{Casrouge:2000be,Hsieh2004267,Hsieh:2006sy,Pacholczyk:2007fs,Pacholczyk:2006rp}). However, this simple model is called into question by the empirical evidence (see e.g., \citealt{Naumov:2003ys,Pewe:2004zr})   suggesting  heavy right tails of the  clonal size distributions.  To account for this heterogeneity, the homogenous model has been  expanded  with  a  variety of mixture models, typically under the  assumption of  the Poisson-distribution of  the TCR clones. These so-called  {\em Poisson abundance mixture models}  \citep{ess06} assume  that each TCR variant (i.e., each clone family)  is sampled according to  the Poisson  distribution with a specific sampling rate, itself varying according to a prescribed parametric (mixing) distribution e.g., exponential, gamma, or lognormal (\citealt{Ord86,Sepulveda:2009ly,Bulmer74}).   A recent  detailed comparative study  of \citet{Sepulveda:2009ly}  identified one of such   models,   the  Poisson-lognormal mixture (PLN), as   particularly well suited for  modeling  clonal diversity.  The special  appeal of the PLN  is in that  it may be naturally extended to multivariate  setting,  allowing therefore  for the simultaneous analysis of abundance patterns of several repertoires \citep{engen2002analyzing}. 
In particular,  the bivariate extension of the PLN model may be used to derive a class of pairwise dissimilarity measures between repertoires and  to construct tree-based hierarchy relating various  TCR repertoires.

The purpose of this article is to describe  a new general approach to analyzing and comparing TCR-type  data arriving from multiple repertoires. The approach we develop here  relies on the TCR repertoires dissimilarities analysis where the  appropriate  tree-distance measures are derived under  a  simple, easily testable  and interpretable model of clonal abundance based on the parametric bivariate Poisson-lognormal  distribution (BPLN).  By means of examples derived from  real TCR data, we argue  that under BPLN both the moment-based and the information-based parametric measures of dissimilarity  yield  consistent and biologically  meaningful   results. 
 The paper is organized as follows. In the next section (Section~\ref{sec:pma}) we give a brief overview of the Poisson abundance models both in univariate and  multivariate (bivariate) settings. In Section~3 we discuss one popular method for deriving dissimilarity measures which is particularly relevant for TCR data studies and provide the formal definitions of the four different   measures considered in the paper. 
  In Section~4 we present the application of our method to TCR data obtained from the populations of naive and so-called regulatory T-cell receptors in healthy and immune-deficient mice. This data set was  described in detail and analyzed by different methods in \citet{Pacholczyk:2006rp}.  We re-analyze it using clustering algorithms derived  under both non-parametric and BPLN  models and compare the results.  In Section~5 we provide a concise summary of our findings  and offer some concluding remarks.  Some elementary derivations related to the entropy function are provided for readers convenience in the appendix.

\section{Poisson Models of  Abundance}\label{sec:pma}
 Poisson abundance models arrive naturally in   the biodiversity  studies  if we  assume (see, e.g., \citealt{ess06})  that the species sampling is done by a ``continuous type of effort"   i.e., data is recorded  as arriving from a mixture of Poisson processes in time interval from $0$ to $T$ (in what follows we take $T=1$).  This type of model approach can be traced back to Fisher, Corbet and  Williams \citep{Fisher43}.  Consider  $M$ species labeled  from $1$ to $M$. Individuals of the $i$-th species arrive in the sample according to a Poisson process with a discovery rate $\la_i$. If the detectability of individuals can be assumed to be equal across all species (which is typically a case in TCR repertoires  analysis), then the rates can be interpreted as species abundances \citep{NAYAK:1991hq}.  In this sampling scheme, the sample size $n$ (the number of individuals observed in the experiment) is a random variable. Since the conditional frequencies  follow a multinomial distribution  with class total $n$ and class probabilities given by relative frequencies $\la_i/\sum_{k=1}^M\la_k$,  many estimators are shared in both the continuous-type Poisson models and the discrete-type (multinomial) models where $n$ is assumed to be a constant. We note here that in case of antigen receptors data,  the constant $n$ is  sometimes known (e.g., DNA sequencing data) and sometimes not known (e.g.,  spectratype data, see, \citealt{Kepler:2005qq}). In the latter case the prior distributional form of  $n$ is typically assumed and the posterior model is investigated based on the Bayesian tools (see, e.g., \citealt{Rodrigues:2001nq,LEWINS:1984ug,Barger:2008zo,SOLOW:1994zi} ).  
Since the present paper is motivated by the single--cell DNA sequencing data, we are assuming throughout that $n$ is known. The 
extension of our model to unknown $n$  along the lines of \citet{Rodrigues:2001nq} is reasonably straightforward but not pursued here. 
\subsection{Univariate mixture models}\label{ss:par} 
Since  it has been generally accepted  that the antigen receptor  clonal size distributions have heavy right  tails, to adjust for  the  over-dispersion  the species rates ($\la_1, \la_2, \ldots, \la_M$) are typically  modeled as a random sample from a mixing distribution with density $f (\la; \bt)$, where $\bt$ is a low-dimensional vector of parameters. 
Following the famous paper by Fisher and his colleagues \citep{Fisher43} many researchers have adopted a gamma density as a mixing model. Other parametric models  include among others the log-normal \citep{Bulmer74}, inverse-Gaussian \citep{Ord86}, and generalized inverse-Gaussian \citep{Sichel:1997yt} distributions as well as many others \citep{Sepulveda:2009ly}. 
An obvious advantage of such parametric models is that the inference problem reduces to estimating only a few relatively low-dimensional parameters for which  the traditional estimation procedures can be typically applied. 
For any mixture density $f (\la; \bt)$, define 
$p_\th(k), k = 0, 1, \ldots$  as the probability that any TCR species is observed $k$ times in the sample, that is
\begin{equation}\label{eq:pk} p_\th(k) = \int^\infty_0 [\la^k e^{-\la}/k!]\,f (\la; \bt)d\la\qquad k=0,1,\ldots.\end{equation} Denoting by $f_k$ ($k=1,2,\ldots,n$) the number of receptor species observed exactly $k$-times in the sample we have $E(f_k) = M p_\th(k)$.  Setting $D=\sum_k f_k$ the likelihood function for $M$ and $\bt$ can be written as 
\begin{equation*}\label{eq:lkh01}   
L(M, \bt|\{f_k\}) = \frac{M !}{(M-D)!  \prod_{k\ge 1} (f_k !)} [p_\th(0)]^{M-D} \prod_{k\ge 1}  [p_\th(k)]^{f_k}. 
\end{equation*}
The (unconditional) MLEs for $M$ and $\bt$ and their asymptotic variances are obtained based on the 
above likelihood which as we can see depends on the data only through the observed values of $\{f_k\}$. The likelihood can be factored as \begin{equation*} L(M, \bt|\{f_k\}) = L_b (M, \bt|D)L_c (\bt|\{f_k\},D)\end{equation*} 
where $L_b (M, \bt|D)$ is a likelihood with respect to $D$, a binomial $(M, 1 - p_\th(0))$ variable, and 
$L_c (\bt|\{f_k\},D)$  is a (conditional) multinomial likelihood with respect to 
$\{f_k; k \ge 1\}$ with cell total $D$ and zero-truncated cell probabilities 
$p_\th(k)/[1 - p_\th (0)], k \ge 1$, i.e., 
\begin{equation}\label{eq:lkh2g}   
L_c(\bt|\{f_k\},D) = \frac{D !}{ \prod_{k\ge 1} (f_k !)}  \prod_{k\ge 1}  \left[\frac{p_\th(k)}{1-p_\th(0)}\right]^{f_k}. 
\end{equation} The MLE obtained from this likelihood  can  be regarded as a (conditional) {\em empirical 
Bayes estimator} if we think of the mixing distribution as a prior distribution having unknown 
parameters that must be estimated. See,   e.g., \citet{Rodrigues:2001nq} for further reference.
\subsection{Extension to bivariate models}
As we shall see in the next section, it is of interest to also consider  multivariate models of abundance. For the purpose of our discussions below we focus on  the bivariate models  but the modifications for higher dimensions are rather straightforward.    For simplicity assume that we have the same  $M$ species in both populations.   In  direct analogy with the  notation of  the previous section, define now $p_\th(k,l)$   to be the probability that any TCR species (i.e., a TCR clone) is present $k$ times in the sample from the first population (repertoire) and $l$ times in the  sample from the second one. Let  $f_{k,l}$ be the empirical count and  set now $D=\sum_{k,l\ge 0} f_{k,l}$ (assuming $f_{0,0}=0$). Let  $f(\la_1,\la_2,\bt)$ be the bivariate mixture distribution. The likelihood formulae from previous section extends to the bivariate case as 
\begin{equation*}\label{eq:lkh1}   
L(M, \bt|\{f_{k,l}\}) = \frac{M !}{(M-D)!  \prod_{k,l\ge 0} (f_{k,l} !)} [p_\th(0,0)]^{M-D} \prod_{k,l\ge 0}  [p_\th(k,l)]^{f_{k,l}}. 
\end{equation*}
where 
\begin{equation}\label{eq:pkl} p_\th(k,l) = \int^\infty_0 [\la_1^k e^{-\la_1}\la_2^l e^{-\la_2}/(k!\,l!)]\,f (\la_1,\la_2; \bt)d\la_1d\la_2\qquad k,l=0,1,\ldots.\end{equation}
Note  that, as before, $E(f_{k,l}) = M p_\th(k,l)$. The likelihood function for $M$ and $\bt$ can be again  factored as \begin{equation*} L(M, \bt|\{f_{k,l}\}) = L_b (M, \bt|D)L_c (\bt|\{f_{k,l}\},D)\end{equation*} 
where, in obvious analogy with \eqref{eq:lkh2g}, $L_b (M, \bt|D)$ is now a likelihood with respect to $D$,  a binomial variable with parameters $(M, 1 - p_\th(0,0))$  and 
$L_c (\bt|\{f_{k,l}\},D)$  is a (conditional) multinomial likelihood with respect to 
$\{f_{k,l}, k+l>0\}$ with cell total $D$ and  the bivariate, zero-truncated  cell probabilities \break
{$\{p_\th(k,l)/[1 - p_\th (0,0)]\}_{k,l}, k+l>0$}, i.e., 
\begin{equation}\label{eq:lkh2}   
L_c(\bt|\{f_{k,l}\},D) = \frac{D !}{ \prod_{k,l\ge 0} f_{k,l} !}  \prod_{k\ge 1}  \left[\frac{p_\th(k,l)}{1-p_\th(0,0)}\right]^{f_{k,l}}. 
\end{equation}

\section{Diversity Analysis and Clustering} 

When studying evolution of TCR species it is of interest to compare their  diversity, by which we mean herein (cf. Section~1) the clonal size distribution $\{p_\th(k)\}$ and the species number $M$.  Such   repertoire diversity comparisons  are   of great interest for instance in clinical studies,  where the quantities of interest are the ``divergences" of multiple observed TCR repertoires from some control or asymptotic one. The individual  repertoires of antigen receptors can be then characterized in terms of their divergence  from the control  \citep{Chen:2003in,Komatsu:2009zr,Pacholczyk:2007fs,Pacholczyk:2006rp}. Under our  definition the TCR repertoire diversity is  completely determined by the  parameters $(M,\bt)$. This agrees with the original concept of ``species  diversity"  known from the field of ecology where the term itself relates both to the number of species (richness) and to their apportionment within the sequence (evenness or equitability, see \citealt{sheldon1969equitability}). A sensible  method  of comparing  diversity of multiple repertoires simultaneously  is based on  a concept of (pairwise) {\em diversity dissimilarity measure} and the  {\em hierarchical clustering}  induced by it. The hierarchical clustering which we discuss in more detail below, is one of many modern methods  of analyzing patterns in high-dimensional data on the grounds of the   so-called unsupervised statistical learning theory (cf., e.g., \citealt{HTF01}), a very dynamically developing area of modern statistics.  

\subsection{Diversity dissimilarity measures}
Assume  that the overall ``similarity"  between a pair of TCR repertoires with respective clonal abundance distributions $p$ and $q$ is quantified by some  non-negative function $Q(p,q)$  referred to  as the  {\em  similarity index} or {\em similarity measure}. 
Since typically the samples from the  joined distribution (frequency) of abundance are not available in data collected from TCR repertoires,  some of the  crude similarity indices are  based simply on the joined TCR species presence/absence data, i.e.,   the number of TCR species  shared by two samples and the number of species unique to each of them \citep[see discussion in][]{LL98}. Examples of such indices are 
 the classical Jaccard index  and the closely related S{\o}rensen index,  the two
oldest and most widely used similarity indices in ecological biodiversity studies \citep{Magurran:2005by}.  One of the advantages of the  Poisson mixture model described in the previous section,  is that it allows for defining  meaningful  indices  incorporating  pairwise comparisons of the TCR species based on joined distribution of abundance.  As  representative examples of such indices we consider  here  the Morisita-Horn index  ($\D_{MH}$)  \citep{Magurran:2005by},  the mutual information criterion ($\D_{MI}$) which is a special case of a Kullback-Leibler divergence \citep[see, e.g., ][]{koski2001hidden},  and the overlap index ($\D_{OV}$) introduced by \citet{smith1996estimator}. All of these indices give rise to  the  corresponding  measures of dissimilarity concentrated on the unit interval with  the perfect correlation (or complete overlap) between  the frequency   distributions   yielding the value zero. Indeed, they   are all seen as special cases of the following general construction. For any bivariate  probability distribution $p_\th$  with corresponding marginal distributions $p_\th^{(1)}$ and  
$p_\th^{(2)}$ consider  a similarity index   $\Q(p_\th^{(1)},p_\th^{(2)})$  satisfying 
\begin{equation}\label{eq:g_cond}
0\le  \Q(p_\th^{(1)},p_\th^{(2)})\le  \frac{\Q(p_\th^{(1)},p_\th^{(1)})+ \Q(p_\th^{(2)},p_\th^{(2)})}{2}
\end{equation} with the  right bound attained when $p_\th^{(1)}=p_\th^{(2)}$.
Then  the corresponding (normalized)  $\Q$-induced measure of dissimilarity between the pair $(p_\th^{(1)},p_\th^{(2)}) $   may be defined as 
\begin{equation}\label{eq:diss} \D(p_\th^{(1)},p_\th^{(2)})=1-\frac{2\,\Q(p_\th^{(1)},p_\th^{(2)})}{\Q(p_\th^{(1)},p_\th^{(1)})+\Q(p_\th^{(2)},p_\th^{(2)})}.  
\end{equation}

To obtain  the {\em Morisita-Horn dissimilarity index} ($\d_{MH}$) we take in \eqref{eq:diss} $\Q=\Q_{MH}$ 
\begin{equation}\label{eq:mh}
\Q_{MH}(p_\th^{(1)},p_\th^{(2)})=\sum_{k,l\ge 1} k\,l\, p_\th(k,l).
\end{equation} which obviously satisfies \eqref{eq:g_cond} for any non-negative random variables. 
A closely related {\em correlation-based dissimilarity index} $\D_\rho$ is obtained when 
we take  $\Q=\Q_\rho$  with 

\begin{equation}\label{eq:rho}
\Q_{\rho}(p_\th^{(1)},p_\th^{(2)})=|\sum_{k,l\ge 0} \tilde{k}\tilde{l}\, p_\th(k,l)|.
\end{equation}
where $\tilde{k}=(k-m_1)/s_1$ and $\tilde{l}=(l-m_2)/s_2$  and $m_i$ and $s_i$ are, respectively, mean and standard deviation of $p_\th^{(i)}$, $i=1,2$.  In this case the  inequality  \eqref{eq:g_cond} simply asserts that $0\le \Q_\rho\le 1$.

A popular  dissimilarity measure  which we shall denote here by $\D_{OV}$   is obtained by averaging conditional probabilities of individual receptor species presence in both samples, given its  presence in  one. The measure was introduced by \citet{smith1996estimator} to quantify 'overlap' between repertoires and is obtained by  taking the following similarity index.
\begin{equation}\label{eq:ov}\Q_{OV}(p_\th^{(1)},p_\th^{(2)})=\frac{\sum_{k,l>0} p_\th(k,l)}{2}\left( \frac{1}{\sum_{k>0}p_\th^{(1)}(k)}+\frac{1}{\sum_{l>0}p_\th^{(2)}(l)} \right)
\end{equation}  which again trivially satisfies \eqref{eq:g_cond}. 

Finally, 
 in order to obtain the {\em mutual information dissimilarity index} ($\D_{MI}$) we take in \eqref{eq:diss} $\Q=\Q_{MI}$  where
\begin{equation}\label{eq:mi}
\Q_{MI}(p_\th^{(1)},p_\th^{(2)})=\sum_{k,l\ge 0} p_\th(k,l) \log\left(\frac{p_\th(k,l)}{p^{(1)}_\th(k)\,p^{(2)}_\th(l)}\right).
\end{equation}
 The fact that the above function satisfies \eqref{eq:g_cond} is shown in the appendix.
Note that all the above  similarity indices $\Q_{MH}, \Q_{\rho}, \Q_{OV}$ and $\Q_{MI}$ (and thus also their corresponding dissimilarity measures) depend on  the underlying mixing distribution parameters $\th$ but not explicitly on  the species number $M$.  This  is  desirable since the  quantity $M$ is typically unknown and difficult to estimate  due to often very severe undersampling of the TCR repertoires (cf., e.g., \citealt{Sepulveda:2009ly}).    In the parametric setting considered here, if needed,  the value of $M$ may be estimated (a-posteriori) by either of the estimates   \begin{align}\label{eq:m1} \hat{M}_1&= D/(1-p_{\th}(0,0))\\
\hat{M}_2&= \sum_{k,l\ge 0, k+l>0} f_{k,l}/p_{\th}(k,l)\label{eq:m2}.\end{align}
whose close numerical agreement  usually indicates a  robust fit of the bivariate parametric mixture model. Note that $\hat{M}_2$ is simply a parametric version of the Horvitz-Thompson estimator \citep{horvitz1952generalization}.

\subsection{Hierarchical clustering}  
For a  given   pairwise dissimilarity measure $\D$ of  TCR repertoires, it is a  standard unsupervised statistical learning approach to  simultaneously compare $N$  repertoires in terms of $\D$ by means of  building  hierarchical clusters which are graphically represented by  {\em dendrogarms}  or   ``tree diagrams".  In such hierarchical clustering procedure the TCR data are not partitioned into a particular cluster in a single step, instead, as a name suggests, a hierarchical structure is produced in which the clusters at each level of the hierarchy are created by merging clusters at the next lower level. The main advantage of hierarchical clustering approach  lies in the fact that no cluster number needs to be specified in advance.   Hierarchical clustering is  performed via  either agglomerative methods, which proceed by series of fusions of the $N$ objects into groups, or divisive methods, which separate objects successively into finer groupings. Agglomerative techniques are more commonly used, and this is the method we consider below for TCR repertoires. For a general introduction to clustering and unsupervised learning,  interested reader is referred  to chapter 14 in the popular  monograph \citet{HTF01}.  
 
\subsection{Poisson-lognormal model}
Whereas there are many possible models of parametric bivariate mixture, the recent studies in \citet{engen2002analyzing} and \citet{Sepulveda:2009ly}  seem to indicate that lognormal mixing  distributions may be often an appropriate choice for TCR repertoires modeling. In that spirit we consider herein a bivariate model based on log-binormal variates.   
Under the assumption of random sampling, the number
of individuals sampled from a given receptor species with abundance $\la$
 is Poisson distributed with mean $\la$. If we assume that 
 $\ln \la$ is normally distributed with mean $\mu$ and variance $\sigma^2$ among species, then the vector of individuals sampled from all $M$ species  constitutes a sample from the
Poisson lognormal distribution with parameters $\th=(\mu, \sigma^2)$, where $\mu$ and $\sigma^2$ are the mean and variance of the log abundances.  The corresponding mass function is  of the general form \eqref{eq:pk} and may be  written as 
\begin{equation}\label{eq:pkl2} p(k; \mu,\sigma^2)=\int^\infty_{-\infty} g_k(\mu,\sigma,u) \phi(u)du\end{equation} where $\phi(\cdot)$ is a standard normal density function and 
\begin{equation*}\label{eq:poi}
g_k(\mu,\sigma,u)=\frac{\exp[u\sigma k+\mu k +e^{-(u\sigma+\mu)}]}{k!},\quad k\ge 0
\end{equation*} is the  re-parametrized Poisson distribution. 
Similarly, when we consider pairs of counts of individual receptors    from two different repertoires  we may think of them as a random sample (of size $M$) from  the bivariate Poisson-lognormal distribution (BPLN) with probability mass function given as in  \eqref{eq:pkl}. 
That is, we assume that the log abundances among species have the binormal distribution
with parameters $(\mu_1, \mu_2,\sigma_1, \sigma_2, \rho)$. For a general detailed description of  the  multivariate Poisson distributions, see \citet{aitchison1989multivariate}.    Let $\phi(u,v; \rho)$ denote the  normal bivariate density with correlation coefficient $\rho$, zero means and unite variances. The distribution of BPLN is given in terms of the bivariate probability mass function $p_\th(k,l)=p(k,l; \mu_1,\mu_2,\sigma_1^2,\sigma_2^2,\rho)$ for $k,l\ge 0$ where  
\begin{equation}\label{eq:pmf}
p(k,l;\mu_1,\mu_2,\sigma_1^2,\sigma_2^2,\rho)= \int_{-\infty}^\infty\int_{-\infty}^\infty g_k(\mu_1,\sigma_1,u) g_l(\mu_2,\sigma_2,v) \phi(u,v; \rho)du\,dv.
\end{equation} 
From the above formula it follows in particular that  both   marginals  of BPLN are the  univariate   Poisson-lognormal distributions  \eqref{eq:pkl2} with respective parameters $(\mu_i,\sigma_i^2)$ $(i=1,2)$. 
Since $M$  is usually unknown, when fitting the model we only consider the 
number of individuals for the observed receptor species and thus  the distribution of the number
of observed individual receptors   follows the zero-truncated BPLN distribution with probability mass function 
\begin{align}\label{eq:pkl2_t}
\frac{p(k,l;\mu_1,\mu_2,\sigma_1^2,\sigma_2^2,\rho)}{1-p(0,0;\mu_1,\mu_2,\sigma_1^2,\sigma_2^2,\rho)}.
\end{align}

The maximum-likelihood estimators (MLEs)
of the parameters of this distribution were discussed e.g., in \citet{Bulmer74}
 and more recently in \citet{karlis2003algorithm} and  \citet{engen2002analyzing}. The latter approach is  conveniently  implemented in the freely available R package {\em poilog} \citep{Rcite} which we have  used  in the current paper  to perform all the  necessary parameter fitting. In our setting,  the  model parameters were  calculated from the multinomial conditional likelihood function \eqref{eq:lkh2} where the truncated probability quantity $p_\th(k,l)/(1-p_\th(0,0))$ is given by \eqref{eq:pkl2_t}.

 Under the assumed BPLN model the measures of dissimilarity may be computed either directly or by 
 Monte-Carlo approximations. Denoting the means of the BPLN marginals by \begin{equation}\label{eq:alpha} \alpha_i=\exp[\mu_i+\sigma_i^2/2] \quad i=1,2,\end{equation}  the moment-based dissimilarity measures $\D_{MH}$ and $\D_{\rho}$ are given by 

\begin{align}\D_{MH}=1-\frac{2\alpha_1\alpha_2\,\exp[\rho\sigma_1\sigma_2]}{\alpha_1(1+\alpha_1\exp[\sigma_1^2])+\alpha_2(1+\alpha_2\exp[\sigma_2^2])}\label{eq:p_mh}\\
\D_{\rho}=1-\frac{2\alpha_1\alpha_2\,(\exp[\rho\sigma_1\sigma_2]-1)}{
\sqrt{\alpha_1(1+\alpha_1(\exp[\sigma_1^2]-1))}\sqrt{\alpha_2(1+\alpha_2(\exp[\sigma_2^2])-1)}\label{eq:p_rho}}
	\end{align} where in \eqref{eq:p_rho} the quantities 
	\begin{equation}\label{eq:var}  \alpha_i(1+\alpha_i(\exp[\sigma_i^2]-1))\quad  i=1,2
	\end{equation} are seen to be the marginal variances of the BPLN distribution. The formulae (\ref{eq:p_mh}-\ref{eq:p_rho}) provide for a  convenient way of estimating the dissimilarities $\D_{MH}$ and $\D_{\rho}$  from the  data simply by replacing the unknown distribution parameters by their sample estimates calculated via maximum likelihood, for instance, by  using the numerical algorithms implemented in the ``poilog" R-package.
	
	Unfortunately, due to the fact that the mass function  \eqref{eq:pmf} is not available in a closed form, there are no similar formulae for  the indices  $\D_{MI}, \D_{OV}$. These indices are  typically  approximated by the  Monte-Carlo-based bootstrap  procedures  which  are also used to derive confidence intervals and standard errors of the parameter estimates in the model.  The general justification for such derivations is provided e.g., in \citet{GRC09} and \citet{rem2}. The reader is referred to \citet{efron1997introduction} for a general introduction to the theory and practice of statistical bootstrap which is the computational technique we rely upon heavily in our modeling approach and the data analysis described below.

\section{Application to TCR Repertoire Data}

In this section we illustrate the parametric inference on T-cell
receptor data based on BPLN model and the associated pairwise
dissimilarity measures by analyzing diversity of the repertoires of
T-cell receptors derived from two types of genetically-engineered  (TCR-mini) mice: the ``wild" type
with genetically restricted TCR repertoire and unaltered  repertoire of
self antigens bound to class II MHC, and the ``Ep" (B63VJEp) mice
that in addition to restricted TCR repertoire also express only single,
covalently linked to MHC Ep peptide \citep{Pacholczyk:2006rp}.

\subsection{Data description and processing}
 The  description of  the  transgenic   ``TCRmini" mice  modifications was  already  detailed  in the recent work \citet{Pacholczyk:2007fs}. 
Herein,  we mention only briefly that the  ``TCRmini" mice  is a new generation of  TCR transgenic mouse  where all T-cells express one pre-specified  TCR$\beta$ chain (specifically, the chain V$\beta$14D$\beta$2J$\beta$2.6), and the unique  TCR$\alpha$ mini-locus. This mini-locus allows only for restricted rearrangements of  a single V$\alpha$2.9 segment to one of the two J$\alpha$ (J$\alpha$26 and J$\alpha$2) segments.  These mice have no other loci that encode TCR$\alpha$ chains and therefore their  entire diversity of TCRs is derived from the artificially introduced TCR$\alpha$ mini locus, resulting in greatly altered  TCRs abundance.

\begin{table}
\centering\begin{tabular}[t]{ l | c c c c |c c  c c}
\multicolumn{4}{c}{Ep}&\multicolumn{4}{c}{Wt}\\
&TN1& TR1& TN2& TR2& TN1& TR1& TN2& TR2\\\hline\hline
$\hat{\mu}$ &-4.54 &-3.90 &-3.58 &-3.80 &-3.70 &-2.97 &-3.55 &-2.90\\
Lo	  &-5.37 &-4.81 &-4.59 &-4.71 &-4.86 &-3.77 &-4.46 &-3.81 \\
Up&  -3.31 &-2.43 &-2.40 &-2.54 &-2.59 &-2.13 &-2.38 &-1.87 \\
\hline\hline
$\hat{\sigma}^2$	&2.03 &1.93 &1.66 &1.70 &1.99 &1.48 &1.84 &1.31\\
Lo	&1.60 &1.52 &1.17 &1.21 &1.61 &1.15 &1.46 &0.90\\
Up &2.36 &2.29 &1.99 &2.02 &2.47 &1.74 &2.15 &1.70\\
\end{tabular}\caption{Maximum likelihood estimates of the means and variances of the Ep (TCR-restricted) and Wt (wild-type) mice repertoires along with  bias corrected 95\% confidence bounds generated via parametric bootstrap.}\label{tab:1}
\end{table}

For the purpose of testing our statistical model,   two subpopulations of CD4+ T-cells (i.e., T-cells expressing a surface marker CD4) were collected representing, respectively,  regulatory (TR) and naive (TN) T-cells (where the TR cells are defined as those expressing  the additional marker Foxp3). In addition, these two subpopulations of CD4+ T-cells were isolated either from (1) the peripheral lymph nodes or (2) from the thymus, giving us a total of {\em eight} TCR populations differing by the animal type, T-cell type, and tissue location.  In the thymus, gross of CD4+ T-cells undergo development  and in the lymph nodes CD4+ T-cells are retained unless activated by specific antigen  and therefore two markedly different patterns of clonal abundance  in these organs are generally  expected.  
Additionally, because in the ``Ep" mice  express only single class II MHC/peptide complex, the diversity of CDR3 region of TCR$\alpha$ chain is drastically reduced in comparison with TCR-mini ``wild" mice.  For these reasons    the  dataset  is uniquely suitable for testing   the BPLN  model. 

The TCR data from both types of mice was collected as follows (for more details on a similar data harvesting procedure, see  also e.g., \citet{War_Nel09,Freeman:2009jh}). Using specific fluorescent reagents,  populations of T-cells from different organs were separated  into individual wells on 96-well plates and amplified in their unique CDR3 regions of their TCR$\alpha$ chain using single-cell RT-PCR (see, e.g., \citealt{Kuczma:2009nt}). Following this amplification, the CDR3 regions were sequenced and analyzed providing the distribution of these regions in native subpopulation(s) of T-lymphocytes. 
This type of  procedure  has  been widely considered to be  one of the most reliable methods to harvest T-cell repertoires \citep{Luczynski:2007wd}. Single T-cells can be separated from cell suspension or be isolated from tissue sections. Both the $\beta$ and the $\alpha$ chains can be amplified and sequenced to provide unambiguous identification of T-cell clones. This method avoids the problems of skewed PCR amplification and varying TCR mRNA expression in different cells. Its obvious drawback is the under-sampling issue alluded to already in Section~1: a very large number of cells need to be analyzed to ensure detection of rare clones and to provide a global representation of the T-cell repertoire. We note that with the availability of the next generation sequencing technology \citep{Wong:2007ph} the  large number  of single cell RT-PCR  could be replaced with the  high throughput PCR from heterogeneous population of T-cells. However, at its current stage the  technology is not yet recommended for repertoires analysis  due to  difficulties of  matching specific TCR$\alpha$ and TCR$\beta$ chains when amplified simultaneously. In addition, there is also a high risk of counts bias due to the skewed  amplification process in high throughput data which tends to  over-express the most dominant DNA sequences and under-express (or even remove) the rare ones.

\subsection{Analysis under BPLN model} 
In order to assess the usefulness of a proposed parametric method of TCR data modeling, we have first generated the BPLN model-based estimates of the dissimilarity matrix for the eight TCR repertoires using separately each of the four dissimilarity measures  $\D$ of the general form \eqref{eq:diss} described in Section~3. For the moment-based  dissimilarities  $\D_{MH}$ and $\D_\rho$ we have used the explicit formulae \eqref{eq:p_mh} and \eqref{eq:p_rho} whereas for the remaining two measures $\D_{OV}$ and $\D_{MI}$ we have directly approximated  the quantities $p(k,l; \mu_1, \mu_2, \sigma_1^2,\sigma_2^2,\rho)$  given by \eqref{eq:pmf} and subsequently used the parametric bootstrap  procedure to produce  the empirical approximations of dissimilarities.  In all cases   the BPLN distribution parameters were estimated by the   maximum likelihood estimators (MLEs) computed  by maximizing the conditional multinomial likelihood function \eqref{eq:lkh2} based on the zero-truncated probabilities \eqref{eq:pkl2_t}. In principle, one could use the conditional  likelihood of multivariate   Poisson-lognormal distribution directly to estimate all of the parameters simultaneously, however,  due to the complicated form of the resulting mixture probabilities, we have deemed that approach to be  too unreliable numerically. On the other hand,  the iterative bivariate model fitting (fitting one BPLN model at a time, conditionally on the remaining repertoires and iterating until convergence)   was  seen to be a reasonable  fast and  numerically stable procedure yielding  a set of estimates  consistent  with marginally MLE-fitted parameters,  regardless of the order in which conditioning was performed.   

The results of the conditional MLE procedure are partially summarized  in Table~\ref{tab:1} where the estimates of  the $\mu$ and $\sigma^2$ parameters for Poisson-lognormal abundance distributions for  the  eight repertoires are reported along with the corresponding confidence intervals obtained via the parametric bootstrap bias-corrected percentile method (see e.g., \citealt{rem2} for details on bootstrap-based interval estimation). Note that the parameters $\mu$ and $\sigma^2$  are  related to the marginal estimates of means and variances of the Poisson-lognormal variates by the formulae \eqref{eq:alpha} and \eqref{eq:var}, respectively.  In order to conserve space, the estimates of the BPLN  correlation coefficients are not shown as they are similar in relative values to   the moment-based dissimilarities   summarized below in Figure~\ref{fig:parden}. We note that the marginal values of the repertoire-specific parameters in both types of repertoires (wild-type and Ep) were found to be  of similar magnitude (with estimated  values of $\mu$ parameters   between -4.5 and -2.9 and $\sigma^2$ parameters  between 1.3 and  2.)  Overall, the numerical values of the parameters   indicated smaller Poisson-lognormal means for the restricted-repertoire mice as compared with  the wild-type. Additionally,   the  naive T-cell repertoires  generally seemed to have   smaller means and larger variances of the mixing log-normal  distributions then the regulatory T-cell repertoires.  

\begin{figure}[htbp] 
   \centering
   \includegraphics[width=5in]{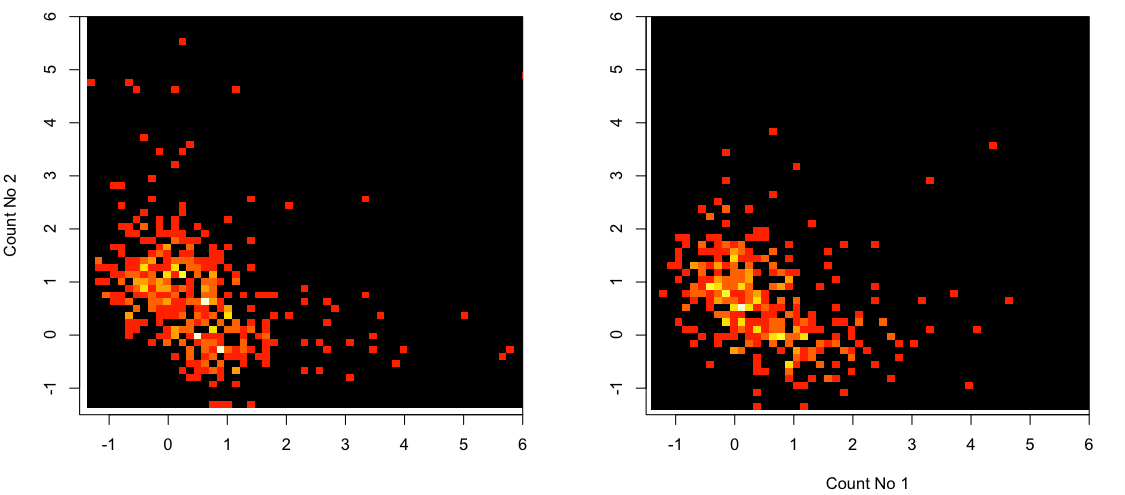} 
\caption{{\bf Model generated vs observed pairs of frequency data.} {\em Left panel:} kernel-density-smoothed heatmap of  joined frequency data of lymph nodes naive ($x$-axis) and thymus regulatory ($y$-axis) TCR repertoires  in wild mice (Wt.TN1 and Wt.TR2 ). {\em Right panel:} kernel-density-smoothed heatmap for the same-size sample simulated from the  distribution of BPLN random variable fitted to the data. Increased brightness indicates higher frequency. }
   \label{fig:heat}
\end{figure}

The goodness of fit statistics were calculated for the bivariate marginal fit via  the  bootstrap distribution of  the conditional likelihood statistic \eqref{eq:lkh2}.  In all cases the differences between the bivariate data and fitted models were not-significant (all $p$-values $<$.05) as measured by the bootstrap tests, indicating reasonable good fit of the parametric distributions to the (zero-truncated) abundance data. 
In addition to the goodness-of-fit  testing, we  have also performed qualitative comparisons  of BPLN model versus data via smoothed heat-map plots \citep{anderes2009local}. One example of such comparison is  provided in Figure~\ref{fig:heat} where the smoothed heat-map illustrates both true and model-generated  bivariate abundance distributions of  wild-type receptor repertoires:  naive TCRs from lymph nodes  and regulatory TCRs from thymus (i.e., Wt TN1 and Wt TR2).

 One of the major advantages of the parametric BPLN model is that in many practical situations of interest it produces, across a number of commonly used dissimilarity indices of the form \eqref{eq:diss}, hierarchical clustering models which are consistent in terms of the final clusters composition. That is, unlike in the non-parametric analysis, for many TCR datasets the  choice of the dissimilarity measure $\D$ is largely irrelevant when modeling repertoires with BPLN model. This is not surprising since our interpretation of the  dissimilarity measure is that it accounts for  differences in diversity  $(\bth,M)$ between pairs of repertoires. Under the condition (which is often satisfied, see Table~\ref{tab:1}) that $\alpha_1\approx \alpha_2$ and $\sigma_1\approx\sigma_2$ one may show that in the BPLN model  the dissimilarity is a monotone function of the parameter $\rho$ and hence measures the correlation between repertoires. This is indeed the case for all indices  $\D$ discussed in Section~3 above.  
  
\begin{figure}[ht]
\centering
\includegraphics[width=5in]{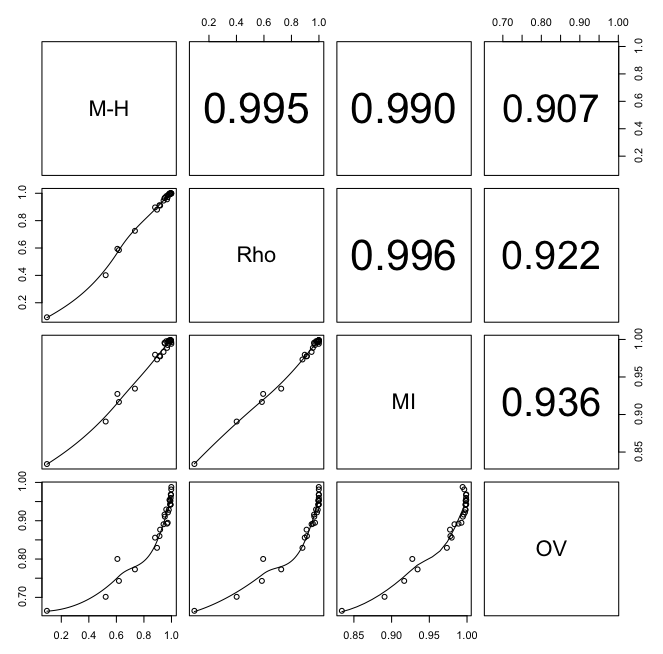} 
\caption{{\bf Pairwise comparison of dissimilarities under BPLN model.}  {\em Lower triangular panels:} pairwise dissimilarities plots obtained  under   BPLN models  fitted to mice data for the four different  $\D$ measures discussed in Section~3.1. Local (loess) regression curves  were added to the plots for better readability.  {\em Upper triangular panels:} $R^2$ values for the corresponding linear regressions.}\label{fig:pairs}
\end{figure}

The results of hierarchical clustering  analysis of the eight mice TCR repertoires under $\D_{MH}$ and $\D_{MI}$ are presented in Figure~\ref{fig:parden}. The figure's left panels (top and bottom)  show the dendrograms obtained by agglomerative hierarchical clustering with a complete link function using $\D_{MH}$ (top) and $\D_{MI}$ (bottom) as the dissimilarity measures. Both dendrograms indicate a very good relative agreement of the cluster hierarchical  structure and the correct final classification of the eight repertoires in terms of the experimental  condition (restricted vs wild-type) as well as the repertoire type (naive or regulatory) and tissue type (thymus vs lymph nodes). Almost identical dendrograms (not shown) were also produced by applying the remaining two measures discussed, namely $\D_\rho$ and $\D_{OV}$. The central  panels in Figure~\ref{fig:parden} illustrate the bootstrap approximations to the distribution of the  Frobenius distance  (see, e.g., \citealt{golub1996matrix} for more on matrix distances) of the dissimilarity matrix  $[\D_{MH}(i,j)]$ (top) and $[\D_{MI}(i,j)]$ (bottom) $(1\le i,j\le 8)$, with 95\% confidence bound marked with vertical lines. The right panels show the dendrograms corresponding to the dissimilarity matrices at the  upper bound of the corresponding 95\% confidence intervals (i.e., right vertical lines of the central panels). The fact that the left and right 
  dendrograms in top and bottom panels  have  identical relative hierarchies indicates  strong robustness of the hierarchical clustering against the fluctuations of both  $\D_{MH}$ and $\D_{MI}$. The  similar    robustness was   also true for $\D_\rho$, and $\D_{OV}$. 
  This  agreement between the dissimilarity matrices entries generated under BPLN model using  four different   dissimilarity measures $\D_{MH}$, $\D_\rho$, $\D_{MI}$ and $\D_{OV}$ is further  illustrated in Figure~\ref{fig:pairs} where the pairwise loess  regressions \citep{cleveland1988locally} of the entries of the dissimilarity measures on each other are presented indicating their monotone  relationship (almost  a linear one between  the first three measures). 
As already indicated, such agreement should be expected if  the data  closely follows the  BPLN model.

As our final analysis for BPLN model, we have also computed the species richness estimates  \eqref{eq:m1} and \eqref{eq:m2} for each pair of repertoires (i.e, each in the total of 28 comparisons) and averaged the result to obtain a pooled estimator of the ratio $D/M$ which was found to be $.09$ with the 95\% confidence interval of (.06, .11).   These  values suggest a much more  severe  under-sampling  of the TCR populations  then the traditional  non-parametric estimator of \citet{good1953population} given by   $f_1/D\approx .25$ (where now  $f_1$ and $D$ are computed from pooled  repertoire data). 

\subsection{Analysis under non-parametric model} 

In order to further examine the results of our parametric hierarchical clustering  analysis of the TCR repertoires,  we have also performed the more traditional, non-parametric hierarchical clustering analysis of the repertoires in which  we have estimated the values of the two dissimilarity  measures $\D_{MH}$ and $\D_{MI}$ with the  non-parametric estimates based directly on the sample data.  Note that $\D_{MH}$ is  particularly convenient to analyze non-parametricaly  as it only requires the relative estimates of the mixed and marginal moments  of order two, which may be calculated  directly from the observed (zero-truncated)  joined abundance. For that reason, the parametric and non-parametric measures $\D_{MH}$ may be directly compared with each other. On the other hand, the information-based measure $\D_{MI}$ may be estimated by means of a recently popularized Chao-Shen estimator \citep{Chao:2003tm} which, in a manner similar to the  ACE and Horwitz-Thomson estimates  \citep[see, e.g.,][] {ess06}, attempts to adjust  explicitly  for the fact of only observing  the truncated  joined distribution.

\begin{figure}[ht]
\centering
\includegraphics[width=5in]{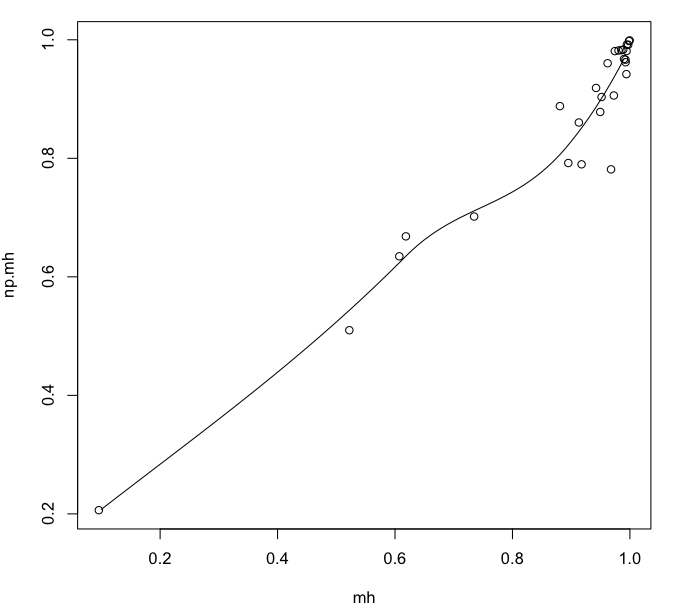} 
\caption{{\bf Morisita-Horn ($\D_{MH}$)  dissimilarities under parametric and non-parametric models.} Scatter plot along with the local (loess) regression curve for  pairwise dissimilarities between eight repertoires computed under parametric (x-axis) and non-parametric (y-axis) models using Morisita-Horn index as given by \eqref{eq:mh}. In the non-parametric case the joined  probabilities $p_\th(k,l)$ are estimated by the corresponding joined frequencies.}\label{fig:mhs}

\end{figure}

\begin{figure}[htbp] 
   \centering
   \includegraphics[width=5in]{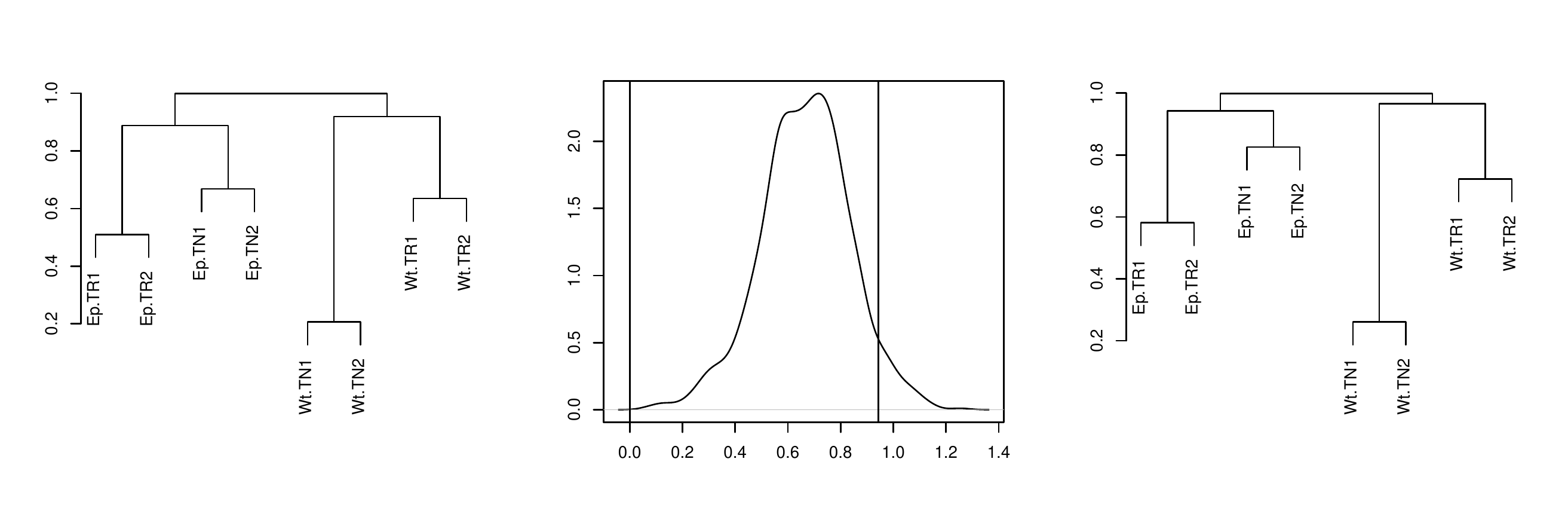}
  \includegraphics[width=5in]{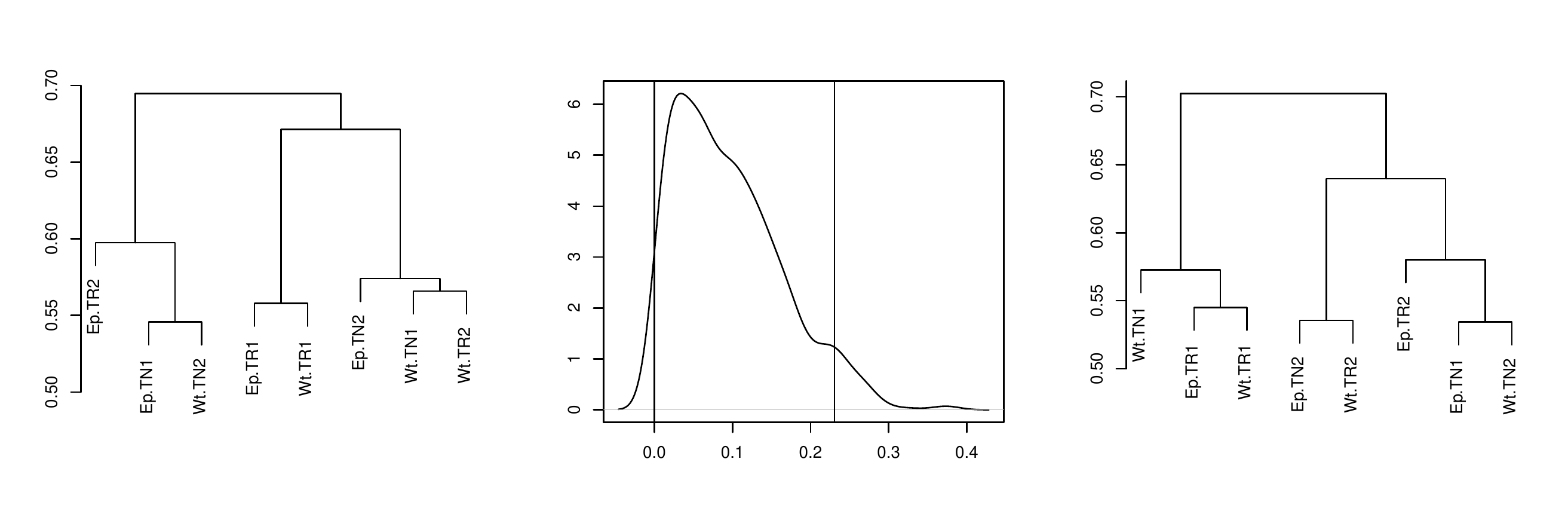} 
\includegraphics[width=5in]{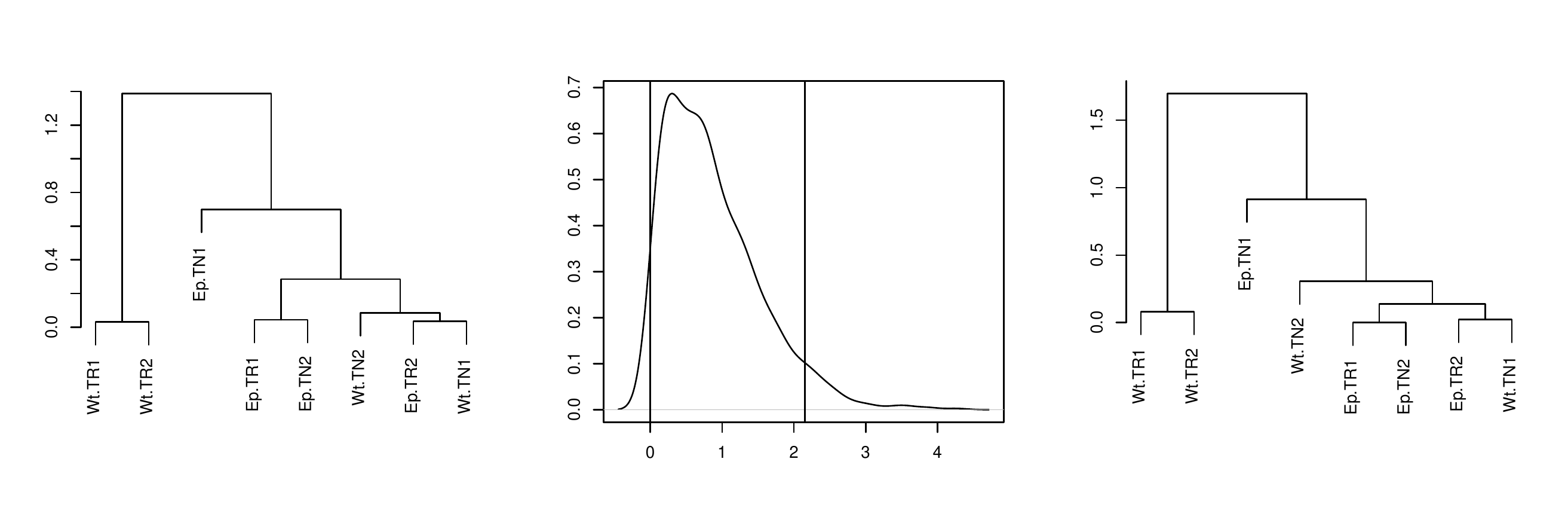}
   \caption{{\bf Repertoire dendrograms and their confidence bounds  under  non-parametric measures of dissimilarity.} Dendrograms under various dissimilarity measures obtained from the non-parametric analogue of the model \eqref{eq:pkl2_t} using agglomerative clustering and the complete link.   In all three panel rows  the most left  dendrograms were obtained using  the point estimates of the dissimilarity matrix calculated from the data,  whereas the most right ones were obtained using upper 95\% confidence bound on the  Frobenius norm distribution of the dissimilarity matrix.    The corresponding bootstrap estimate  of the entire norm distribution is provided in the middle plot as a density estimator, with 95\% bounds marked with vertical lines. {\em Top panels}: hierarchical clusters based on the nonparametric version of the  Morisita-Horn dissimilarity  index \eqref{eq:mh}. 
   {\em  Middle panels:} clusters based  on the non-parametric version of the mutual information dissimilarity \eqref{eq:mi}. {\em  Bottom panels:} clusters based on the values of the Shannon entropy function with no direct pairwise comparisons.    }
   \label{fig:exam}
\end{figure}

 The result of the direct comparison between all the  pairwise estimated $\D_{MH}$ values under the BPLN  and the non-parametric models is presented  as a scatter plot with fitted loess trend-line in Figure~\ref{fig:mhs}.  The plot follows a linear trend indicative of a very close  agreement between non-parametric and parametric $\D_{MH}$ values for our TCR data.  This  apparent almost linear relationship   between the dissimilarities estimated under the two measures  is also confirmed by the  dendrogram computed under non-parametric $\D_{MH}$ (top panel in Figure~\ref{fig:exam}) which gives a stable set of hierarchical clusters almost identical  to those obtained under BPLN model (top panel in Figure~\ref{fig:parden}). 

 In contrast to the parametric case, we found that  the non-parametric analogue of  the mutual information (MI) dissimilarity  \eqref{eq:mi}, based on the coverage adjusted  Chao-Shen entropy estimator \citep{Vu:2007im}, did not agree with the non-parametric Morisita-Horn (MH) dissimilarity and consequently yielded a very different and  biologically uninterpretable TCR clustering which  lacked separation between the wild-type and Ep mice.  
 These differences  between non-parametric MH and MI measures may be clearly seen in the two top panels of  Figure~\ref{fig:exam} where also the lack of stability of the MI-dissimilarity is clearly manifested by the large difference between  dendrograms within the 95\% confidence bound induced by the Frobenius norm of the MI-dissimilarity matrix. Note that this is not the case for the MH dissimilarity, which appears quite stable. 
 
The lowest third panel of Figure~\ref{fig:exam} illustrates  a different  non-parametric dissimilarity analysis we have performed, based on the coverage adjusted estimated values of the Shannon  entropy function for  the eight repertoires (see e.g., \citealt{Vu:2007im}, and formula \eqref{eq:h} in the appendix). The  pairwise dissimilarities were computed as  absolute  differences between the estimated entropy values. Such ``linear" comparisons of the diversity measures across repertoires are often appropriate when the repertoires are assumed to have  similar abundance patterns (see e.g., \citealt{Sepulveda:2009ly}).  However,    as we may see from the plots, for our datasets  the  entropy-based clustering turned out to be only partially satisfactory, as the entropy measure  was only able to clearly separate two repertoires derived from the wild-type regulatory cells but not the remaining ones.   Overall, it  appears that for our dataset the entropy based clustering both via MI and the Shannon entropy performed  poorly whereas the MH clustering was satisfactory and comparable (in fact almost identical)  with our parametric results. These large differences seems to be at least partially due to the fact that, unlike the MH dissimilarity estimate,  the non-parametric entropy estimates are  adjusting for the zero-truncation of the observed distribution and these adjustments in our particular dataset seem to fluctuate widely  from repertoire to repertoire.

\section{Summary and Discussion}
 We have presented  a simple bivariate-Poisson-lognormal  parametric model for  fitting and analyzing TCR repertoires data which may be regarded as a natural multivariate extension of  a Poisson abundance  model with lognormal mixing distribution  which has been applied  in TCR modeling perviously.   As seen from our  example of analyzing TCR naive and regulatory repertoires in the wild-type and TCR-restricted Ep mice, the model provides for a robust and consistent fit to the TCR data allowing for a very detailed yet relatively simple comparison  of multiple repertoires, for instance by means of dissimilarity analysis and hierarchical clustering.  We show   that in our TCR dataset under the  parametric  model the  four popular dissimilarity indices:  (i) Morista-Horn and (ii) correlation indices, (iii) mutual information and (iv) the overlap dissimilarities are monotone functions of each other, leading therefore to the same, biologically meaningful, clustering of the repertoires. In contrast, when applying  to the same data the  non-parametric  estimates of dissimilarities,  the clusterings are seen to behave  erratically  and are highly dependent  on the particular choices of measures with some of them  (e.g.,the  mutual information) yielding biologically implausible clustering. 
 
 Our parametric analysis suggests that in a  typical   experiment based on harvesting sequences from single-cell TCRs, the overall under-sampling  of the TCR population is   much  higher  than in the macroscopic biodiversity studies,  for which many of the statistical tools of species abundance comparison were originally developed.  This  and  the apparent lack of  agreement between the   non-parametric dissimilarity measures  when applied to our, relatively simple, dataset,  seem to indicate that many commonly used non-parametric  biodiversity statistics may perform poorly when applied to severely under-sampled  TCR  repertoires.  The advantage of the parametric approach  and in particular the model  proposed here is that even with very severe data under-sampling it allows for the proper adjustment for the missing abundance information and  estimation of the full set of repertoires features.  The statistical package {\em poilog} available from CRAN archive ({\tt http://cran.r-project.org}) makes the fitting of our  model particularly convenient by providing numerical algorithms for the parameters estimation via the maximum likelihood. Further studies and a larger number of TCR datasets with   more sequences  are needed in order to more comprehensively evaluate the BPLN model and test its ability to discriminate between TCR repertoires in a biologically meaningful way.  
 
\appendix 
\section{Appendix:  Mutual Information Bounds} 
 The fact that the bounds \eqref{eq:g_cond}  hold  for the dissimilarity index \eqref{eq:mi}  follows from the general properties of the Shannon entropy function  which is defined (see e.g., \citealt{koski2001hidden}) for any discrete random vector $X$   with probability distribution $p(x)$  as \begin{equation}\label{eq:h} H(X)=-\sum_x p(x)\log p(x),\end{equation} with the summation is taken  over $x$ values for which $p(x)>0$.
 Extending the definition of the index \eqref{eq:mi} to any pair of  discrete real random variables $X,Y$  with joined distribution $p(x,y)$ and marginals $p(x), p(y)$, we define  their {\em mutual information} as \begin{align}MI(X,Y) &= \sum_{x,y} p(x,y)\log\left(\frac{p(x,y)}{p(x)p(y)}\right)\nonumber\\ &=-\sum_x p(x) \log p(x)-\sum_y p(y) \log p(y)+\sum_{x,y} p(x,y) \log p(x,y)\nonumber\\
 &=H(X)+H(Y)-H(X,Y)\label{eq:mi0}
 \end{align}
 Due to the elementary inequality $\log(x)\le x-1$ valid for any $x>0$ we have that 
$$
-MI(X,Y)=\sum_{x,y} p(x,y)\log\left(\frac{p(x)p(y)}{p(x,y)}\right)\le \sum_{x,y} p(x,y)\left( \frac{p(x)p(y)}{p(x,y)}-1\right)=
- \sum_{x,y} p(x)p(y)+1=0
$$ and therefore 
\begin{equation}\label{eq:mi2}
MI(X,Y)\ge 0
\end{equation}  Note that $MI(X,X)=H(X)$ and therefore (due to symmetry and \eqref{eq:mi0}) to argue upper bound in \eqref{eq:g_cond} it suffices to show  that \begin{equation}\label{eq:mi3} H(X,Y)\ge H(X).\end{equation} This follows easily, since 
$$ H(X,Y)-H(X)=\sum_{x,y} p(x,y) \log p(x,y) -\sum_x p(x) \log p(x)=-\sum_{x,y} p(x,y)\log\left(\frac{p(x,y)}{p(x)} \right)\ge 0.$$
The bounds \eqref{eq:g_cond} for $MI(X,Y)$  follow now from  \eqref{eq:mi2}--\eqref{eq:mi3} and \eqref{eq:mi0}. The measure and  \eqref{eq:mi} is, of course a special case of $MI(X,Y)$.
\bibliographystyle{apalike}
\bibliography{poilog}

\newpage

\begin{figure}[ht]
\centering
\includegraphics[width=5in]{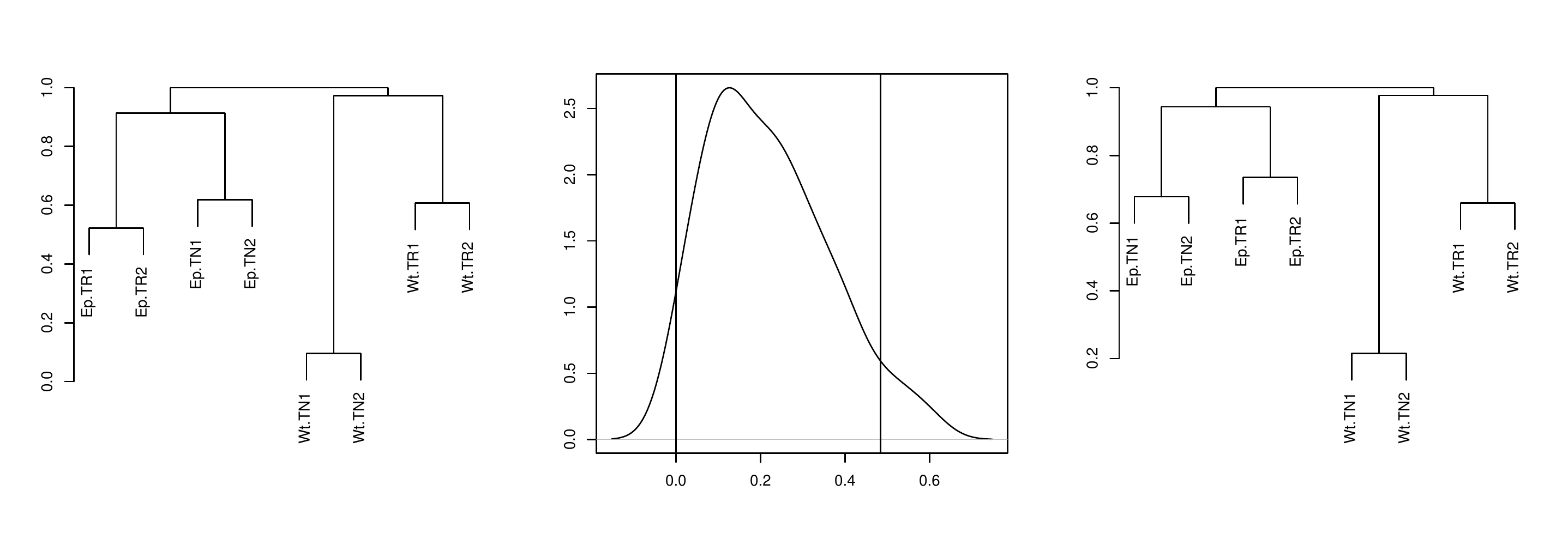} 
\includegraphics[width=5in]{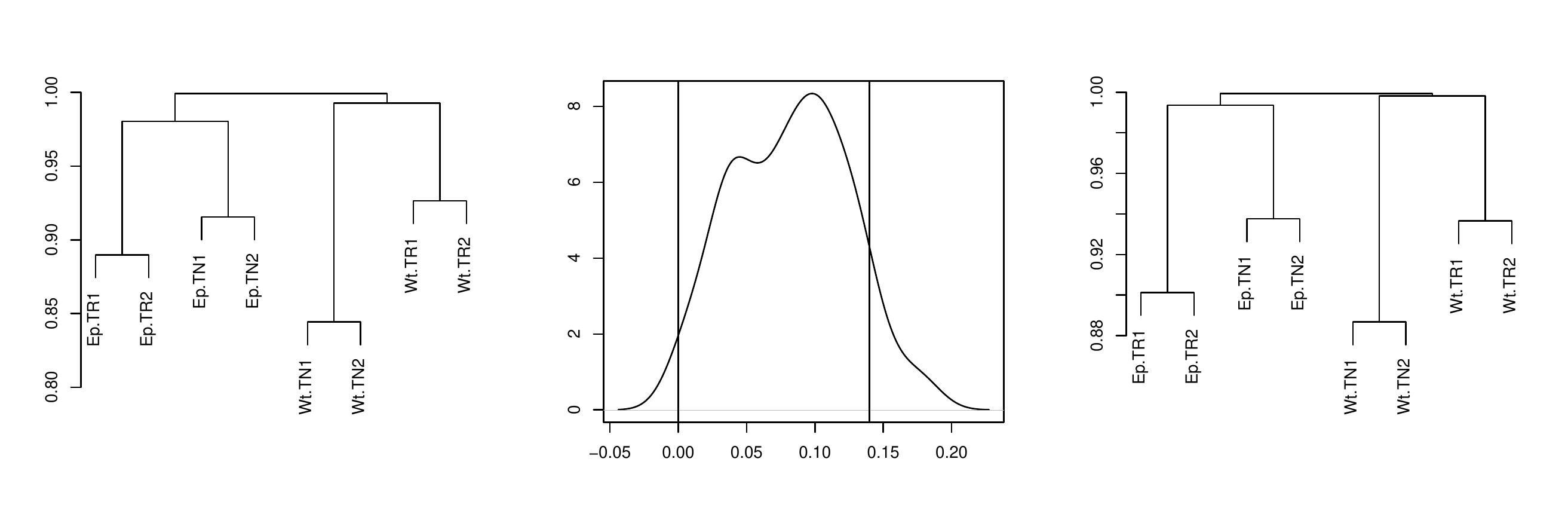}
\caption{{\bf Repertoire dendrograms and their confidence bounds  obtained under  BPLN model.} Dendrograms for hierarchical clustering  and their corresponding confidence intervals obtained using agglomerative clustering and a complete link for  eight repertoires of naive and regulatory TCRs derived from (1) the lymph nodes  and (2)  thymus in wild type and TCR mini mice.{\em Top panel:} (left) clustering using  Morisita-Horn dissimilarity measure $\D_{MH}$ given by \eqref{eq:mh}; (middle) bootstrap estimate of the 95\% confidence interval of the Phrobenius norm of the $\D_{MH}$ dissimilarity matrix; (right) dendrogram corresponding to the upper bound of the 95\% CI $\D_{MH}$ dissimilarity matrix. {\em Bottom panel:}  hierarchical clustering according to  the parametric mutual information  dissimilarity measure $\D_{MI}$ given by \eqref{eq:mi}.}
   \label{fig:parden}
\end{figure}

\end{document}